\documentclass[ pra,aps,superscriptaddress,floatfix, notitlepage,nofootinbib ,preprint,longbibliography]{revtex4-1}
\usepackage[utf8]{inputenc}
\usepackage{amsthm}
\usepackage{amsmath,amssymb}
\usepackage{fullpage}
\usepackage{color}
\usepackage{tikz}
\usetikzlibrary{calc}

\usepackage[title]{appendix}  

\usepackage{thmtools}
\usepackage{thm-restate}

\usepackage[colorlinks=true,citecolor=blue,urlcolor=blue]{hyperref}
\usepackage[capitalize]{cleveref} 

 \usepackage{afterpage}

\newtheorem{result}{Result}
 
\newcommand{\tr}{\operatorname{Tr}}

\newcommand{\bra}[1]{\langle #1 \vert}
\newcommand{\ket}[1]{\vert #1 \rangle}

\newcommand{\Ket}[1]{\left\vert #1 \right\rangle}

\newcommand{\norm}[1]{ \|#1  \|}

\newcommand{\II}[0]{\mathbb{I}}

\newcommand{\Ttot}[0]{  T}
\newcommand{\Tsteps}[0]{ \tau}
\newcommand{\Dt}[0]{   \Delta t}

\newcommand{\LL}{\mathcal{L}}
\newcommand{\BB}{\mathcal{L}}
\newcommand{\HH}{\mathcal{H}}
\newcommand{\TT}{\mathcal{T}}
\newcommand{\XX}{\mathcal{X}}
\newcommand{\MM}{\mathcal{M}}
\newcommand{\CC}{\mathcal{C}}
\newcommand{\WW}{\mathcal{W}}

\tikzset{
	box/.pic = {
		\draw[fill=white] (-1,-0.5) rectangle (01,.5);
	}
}

\tikzset{
	peps/.pic = {
		\draw (-0.5,0)--(0.5,0);
		\draw[fill=white] (-0.25,-0.25) rectangle (0.25,0.25);
		\draw[style=thick] (-.2,.4)--(0,0);
	}
}

\tikzset{
	pepsDOWN/.pic = {
		\draw (-0.5,0)--(0.5,0);
		\draw[fill=white] (-0.25,-0.25) rectangle (0.25,0.25);
		\draw[color=white] (-0.25,-0.25)--(0.25,-0.25);
		\draw[style=thick] (-.2,.4)--(0,0);
	}
}

\tikzset{
	circ/.pic = {
		\draw (0,0)--(0.5,0);
		\draw (0,-.5)--(0,.5);
		\draw[fill=white] (0,0) circle (0.25);
	}
}

\tikzset{
	circno/.pic = {
		\draw (0,0)--(0.5,0);
		\draw[fill=white] (0,0) circle (0.25);
		\node at (0,0) {\tikzpictext};
	}
}

\tikzset{
	circup/.pic = {
		\draw (0,0)--(0.5,0);
		\draw (0,-.5)--(0,0);
		\draw[fill=white] (0,0) circle (0.25);
	}
}
\tikzset{
	circdown/.pic = {
		\draw (0,0)--(0.5,0);
		\draw (0,0)--(0,.5);
		\draw[fill=white] (0,0) circle (0.25);
	}
}
\tikzset{
	unitary4/.pic = {
		\draw (-0.5,0)--(0.5,0);
		\draw (-0.5,1)--(0.5,1);
		\draw (-0.5,2)--(0.5,2);
		\draw (-0.5,3)--(0.5,3);
		\draw[fill=gray] (-0.25,-0.25) rectangle (0.25,3.25);
		\node at (0,1.5) {\tikzpictext};
	}
}

\tikzset{
	unitary5/.pic = {
		\draw (-0.5,0)--(0.5,0);
		\draw (-0.5,1)--(0.5,1);
		\draw (-0.5,3)--(0.5,3);
		\draw (-0.5,4)--(0.5,4);
		\draw[fill=gray] (-0.25,-0.25) rectangle (0.25,4.25);
		\draw[color=white] (-0.25,-0.25)--(0.25,-0.25);
		\node at (0,2) {\tikzpictext};
		
	}
}

\tikzset{
	unitary5white/.pic = {
		\draw (-0.5,0)--(0.5,0);
		\draw (-0.5,1)--(0.5,1);
		\draw (-0.5,3)--(0.5,3);
		\draw (-0.5,4)--(0.5,4);
		\draw[fill=white] (-0.25,-0.25) rectangle (0.25,4.25);
		\draw[color=white] (-0.25,-0.25)--(0.25,-0.25);
		\node at (0,2) {\tikzpictext};
	}
}

\tikzset{
	unitary3white/.pic = {
		\draw (-0.5,0)--(0.5,0);
		\draw (-0.5,2)--(0.5,2);
		\draw[fill=white] (-0.25,-0.25) rectangle (0.25,2.25);
		\node at (0,2) {\tikzpictext};
	}
}

\tikzset{
	greygate/.pic = {
		\draw[fill=gray] ( -.25,-.5 ) rectangle (.25,.5 );
		\node at (0,0) {\tikzpictext};
	}
}

\tikzset{
	whitegate/.pic = {
		\draw[fill=white] ( -.25,-.5 ) rectangle (.25,.5 );
		\node at (0,0) {\tikzpictext};
	}
}

\tikzset{
	whitegateONE/.pic = {
		\draw[fill=white] ( -.25,-.25 ) rectangle (.25,.25 );
		\node at (0,0) {\tikzpictext};
	}
}

\tikzset{
	whitegateLONG/.pic = {
		\draw[fill=white] ( -.25,-1 ) rectangle (.25,1 );
		\node at (0,0) {\tikzpictext};
	}
}

\tikzset{
	mpstensorBIGdash/.pic = {
		\draw (-0.65,0)--(0.65,0);
		\draw [style=dashed] (0,0)--(0,0.75);
		\draw[fill=white] (-0.35,-0.35) rectangle (0.35,0.35);
		\node at (0,0) {\tikzpictext};
	}
}

\begin{document} 	
	\title{A Spacetime Area Law Bound on  Quantum Correlations}

\author{Ilya Kull   }  
\affiliation{Faculty of Physics, University of Vienna, Boltzmanngasse 5, 1090 Vienna, Austria}
\affiliation{Institute for Quantum Optics and Quantum Information (IQOQI),	Austrian Academy of Sciences, Boltzmanngasse 3, 1090 Vienna, Austria}	
\affiliation{Correspondence: ilya.kull@oeaw.ac.at}

\author{Philippe Allard Gu\'erin} 
\affiliation{Faculty of Physics, University of Vienna, Boltzmanngasse 5, 1090 Vienna, Austria}
\affiliation{Institute for Quantum Optics and Quantum Information (IQOQI),	Austrian Academy of Sciences, Boltzmanngasse 3, 1090 Vienna, Austria}	

\author{ \v Caslav Brukner}
\affiliation{Faculty of Physics, University of Vienna, Boltzmanngasse 5, 1090 Vienna, Austria}
\affiliation{Institute for Quantum Optics and Quantum Information (IQOQI),	Austrian Academy of Sciences, Boltzmanngasse 3, 1090 Vienna, Austria}

\begin{abstract}
	Area laws are a far-reaching consequence of the locality of physical interactions, and they are relevant in a range of systems, from black holes to quantum many-body systems. Typically, these laws concern the entanglement entropy or the quantum mutual information of a subsystem \textit{at a single time}. However, when considering information propagating in spacetime, while carried by a physical system with local interactions, it is intuitive to expect  area laws to hold for spacetime regions. In this work, we prove such a law for quantum lattice systems. 
	We consider two agents interacting in disjoint spacetime regions with a spin-lattice system that evolves in time according to a local Hamiltonian. In their respective spacetime regions, the two agents apply quantum instruments to the spins. By considering a purification of the quantum instruments, and analyzing the quantum mutual information between the ancillas used to implement them, we obtain a spacetime area law bound on the amount of correlation between the agents' measurement outcomes. Furthermore, this bound applies both to signaling correlations between the choice of operations on the side of one agent, and the measurement outcomes on the side of the other; as well as to the entanglement they can harvest from the spins by coupling detectors to them.
\end{abstract}

\maketitle

\section*{Introduction}  
How much information is   available to an observer, given access to a \textit{spacetime} region,   about the rest of spacetime? 
Because of the locality of physical interactions, the boundary of the region seems most relevant for information acquisition. Intuitively, we might  expect this information to scale proportionally to the size of the region's boundary.

We shall address this question within the   framework of quantum lattice systems. 
The investigation of quantum information properties of such systems   is of great interest in itself,  as they have profound implications both on the field of condensed matter physics and on quantum computing. 
Furthermore,
 such systems can serve as lattice approximations of relativistic quantum field theories.
The vacuum state of such theories   displays a rich entanglement structure~\cite{Unruh76, SUMMERS1985, Summers1987}  which can  be `harvested', i.e.\     it is possible to produce an entangled state of two initially uncorrelated detectors by making them interact with the field alone~\cite{Martinez2013, Reznik2005,Reznik2003,Olson2011,Sabin2012,Retzker2005}. Fundamental bounds on such entanglement harvesting are also of great interest.

In quantum lattice systems with local interactions, the Lieb-Robinson bound provides a limit on the speed of propagation of information~\cite{Lieb1972}.  
As a result, an effective light-cone structure emerges. In Ref.~\cite{Bravyi2006} it was shown  that an observer,  Bob, effectively cannot detect whether or not   Alice  has manipulated  her part of the system in the past, if he performs measurements outside of her light cone. It was further shown that  
correlations between parts of  the system can not be created by the time evolution in time intervals shorter than that needed for a signal, traveling at the Lieb-Robinson velocity, to reach from one part to the other.
  
 Area law  bounds on  the entanglement entropy are a further consequence of the  locality of interactions.
 First studied in relation to black hole thermodynamics~\cite{Srednicki1993,Callan:1994py,Bombelli1986,Sorkin:2014kta}, they  were observed to hold in ground states of  non-critical quantum lattice systems (see Ref.~\cite{Eisert2010} for a review).  They  deal with the entanglement  structure of specific states, e.g.\ ground states of local Hamiltonians, or  their thermal states~\cite{Wolf2008}.  Yet another property of local Hamiltonians~\cite{Bravyi2006,Marien2016} is that the rate at which they generate entanglement between two regions of the lattice is governed  by an area law.

The  mentioned results    provide bounds on the amount of information that can be shared between regions. In each  bound the amount of information  scales   with the size of the region considered. It is worth noting that  the scaling is not the same in all bounds. In the Lieb-Robinson bound there is a prefactor  which scales with the \textit{volume}  of the smaller of the two regions~\cite{Bravyi2006}, as opposed to  the area law results.    
 Scaling with time appears only in the results about entanglement rates, Refs.~\cite{Bravyi2006,Marien2016}.
 All these results demonstrate how  the locality of interactions restricts the propagation of information between spacetime regions.  The question of its overall \textit{spacetime} scaling, however,         remains open.

To address this question, we adopt an operational definition  for the notion of  propagation of information  between spacetime regions. 
We consider two agents, Alice and Bob, restricted to probing  a time evolving lattice spin system   in disjoint spacetime regions, using  quantum instruments of their choice. We   consider both signaling   and non-signaling  correlations between the settings and the outputs of the   measurement devices used by Alice and Bob.

In this article we prove a spacetime area law bound on correlations   in the presence of local dynamics. 
We show that both the maximal correlation between measurement  outcomes, and the maximal signaling capacity between spacetime  regions,  are bounded by the area of the boundary separating them. Note that this is a co-dimension 1 surface in spacetime whereas the above mentioned results regarding entropy area laws for  spacetime regions~\cite{Srednicki1993,Callan:1994py,Bombelli1986,Sorkin:2014kta} refer to the area of a co-dimension $2$ surface in spacetime. 
We prove this bound for finite-dimensional quantum lattice systems in which time evolution is governed by  local Hamiltonians, and for one-dimensional quantum cellular automata.

 We shall employ  purifications of the instruments used to probe the system. This  will be shown to reduce   the problem of bounding correlations to that of bounding the von Neumann entropy of the reduced state of  ancillas used for the purification of Alice's instruments. 
 Apart from serving as a computational aid,   the purified setup  highlights the affinity  of our setup to that of entanglement harvesting~\cite{Martinez2013, Reznik2005,Reznik2003,Olson2011,Sabin2012,Retzker2005} where such ancillas are called probes or detectors.

 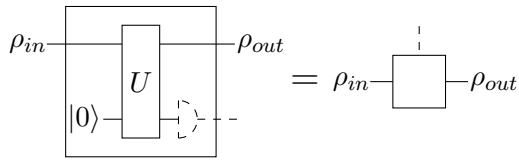
\begin{figure}[h]
 	\begin{tikzpicture}[baseline=-1mm]
 	\pic at (5.2,0) {mpstensorBIGdash};
 	\node at (6.2,0) {$\rho_{out}$};
 	\node at (4.3,0) {$\rho_{in}$};	
 	
 	\node at (3.6,0) {\large{ $=$}};

 	\node at (3.1,.5) {$\rho_{out}$};
 	\node at (0.,.5) {$\rho_{in}$};
 	
 	\draw[fill=white] (.5,-1) rectangle (2.5,1);
 	
 	\draw (0.25,0.5)--(2.75,.5);
 	\draw (1,-0.5)--(2,-.5);
 	\draw [style=dashed] (2.25,-0.5)--(2.85,-.5);	
 	\draw  [fill=white] (1.25,-.75) rectangle (1.75,.75);
 	\node at (1.5,0) { $U$};
 	\node at (.75,-.5) { $\ket{0}$};
 	\draw [style=dashed] (2,-0.75) arc (-90:90:0.25); 
 	\draw [style=dashed] (2,-.25)--(2,-.75);
 	\end{tikzpicture}
 	\caption{ 
 		 		Equivalent representations of a  purified quantum instrument. The recorded measurement outcome is produced by a projective measurement (represented by the dashed half circle) of the ancilla system (initially in the state $\ket{0}$). This measurement can be deferred to a later time. The LHS shows the details of the purification. The RHS representation is an isometry from the input space to the output and ancilla spaces. When consecutive measurements are performed on a system, the resulting state of the ancilla systems at the end of the process admits  a matrix product state representation which is  obtained by concatenating copies of the RHS~\cite{Markiewicz2014}.}
 	\label{fig:POVMMPS1}
 \end{figure}

 The most general  quantum instrument can be implemented by introducing an ancillary quantum system in a pure state (which  we denote by $\ket{0}$); applying a unitary on both system and ancilla; and performing a projective measurement on a part of the composite system to obtain the recorded measurement outcome and the corresponding post-measurement state of the system~\cite{nielsen_chuang_2010}.
 When performing a sequence of measurements, each one involving a fresh ancillary system, such projective measurements can be deferred to the end of the overall process. Up to that point, the physical system and the ancillas undergo a joint unitary evolution.  
 In Ref.~\cite{Markiewicz2014} it was shown that the resulting state of the ancillas (before the projective measurements are made) can be represented  by a matrix product state~\cite{Perez-Garcia2006}. This is illustrated in~\cref{fig:POVMMPS1}. 
 Following the same approach, we will  represent the state of the ancillas after the measurement process as a tensor network state~\cite{Verstraete2008}.

\section*{Results}

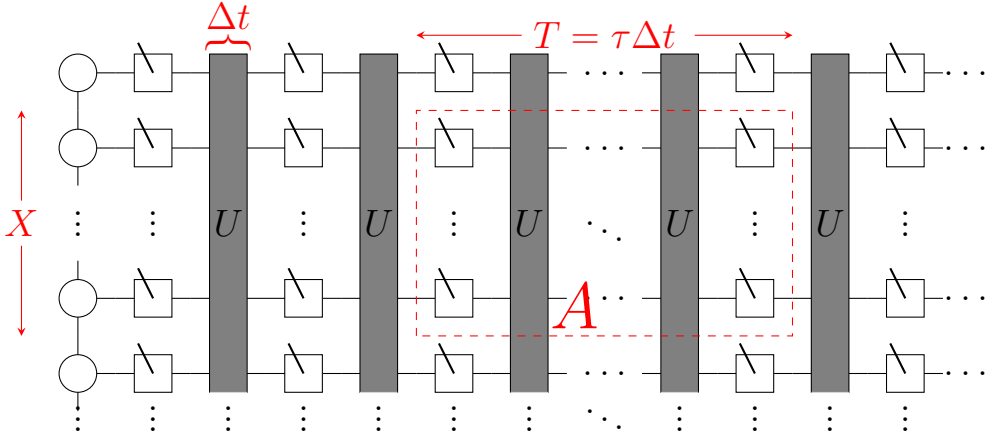
\begin{figure}[h]
 		\begin{tikzpicture}[baseline=-1mm,scale=1, every node/.style={transform shape}]
 		\foreach \j in {0,1,3}    \pic at (0,\j) {circ}; 
 		\pic  at (0,4) {circup};
 		\node at (0,2.1) {\large{$\vdots$}};
 		
 		\foreach \i in {1,3,5, 9,11} \foreach \j in {0,1,3,4}  
 		\pic  at (\i,\j) {peps};

 		\foreach \j in {0,1,3,4} \node at (11.8,\j) {\large{$\ldots$}};	
 		\node at (1,2.1) {\large{$\vdots$}};
 		\node at (3,2.1) {\large{$\vdots$}};
 		\node at (5,2.1) {\large{$\vdots$}};
 		\node at (9,2.1) {\large{$\vdots$}};
 		\node at (11,2.1) {\large{$\vdots$}};
 		\node at (7,2) {\large{$\ddots$}};
 		\node at (7,-.5) {\large{$\ddots$}};
 		\node at (7,3 ) {\large{$\ldots$}};
 		\node at (7,1 ) {\large{$\ldots$}};
 		\node at (7,4 ) {\large{$\ldots$}};
 		\node at (7,0 ) {\large{$\ldots$}};
 		\foreach \i in {0,1,2,3,4,5,6, 8,9,10,11} \foreach \j in {-.5} 
 		\node at (\i,\j) {\large{$\vdots$}};
 		\foreach \i in {2,4,6,8,10}  
 		\pic[pic text = {\large$U$}] at (\i,0) {unitary5};
 		
 		\draw [color=red, style=dashed] (4.5,.5)--(4.5,3.5)--(9.5,3.5)--(9.5,.5)--(4.5,.5);
 		
 		\draw [<-,>=stealth,color=red] (-.75,.5)--(-.75,1.7);
 		\draw [->,>=stealth,color=red] (-.75,2.3)--(-.75,3.5);
 		\node [color=red] at (-.75,2) {\large{$X$}};
 		
 		\draw [<-,>=stealth,color=red] (4.5,4.5)--(5.9 ,4.5);
 		\draw [->,>=stealth,color=red] (8.2,4.5)--(9.5,4.5);
 		\node [color=red] at (7,4.5) {\large{$\Ttot=\Tsteps\Dt$}};
 		
 		\node at (2,4.5) [red,rotate = -90] {\Large { $\{$ }};
 		\node [color=red] at (2, 4.75) { \large  {$\Dt$}};
 		\node [color=red] at (6.6,.9) {{\Huge $A$}};
 		
 		\end{tikzpicture}
 	 	\caption{
 		 		Consecutive local operations followed by Hamiltonian time evolution realize a tensor network state of the ancillas used to perform the operations.  The circles on the left represent the initial state. The white squares represent local instruments   with the upward pointing legs representing the ancillas, as in \cref{fig:POVMMPS1}. The shaded rectangles represent the unitary time evolution operators.~$\mathbf{A}$ is the spacetime  region of interest.~$X$ is the number of spins on which measurements in $\mathbf{A}$ are performed - the spatial extent of $\mathbf{A}$. $\Ttot=\Tsteps\Dt$ is the total   duration of $\mathbf{A}$ in time, where $\Tsteps$ is the number of time steps and $\Dt$ is the duration of time evolution between measurements.}
 	\label{fig:TN}
 \end{figure}

For the sake of clarity   we shall first present the setting of the problem for a system in one spatial dimension. In this case, the graphical representation of the problem is instructive and easy to follow (see \cref{fig:TN}). The generalization to  higher spatial dimensions is straightforward and  our results apply to spin lattices of any spatial dimension.

Consider a `spin' chain, i.e.\ identical $d$-dimensional quantum systems  positioned on a one-dimensional lattice. 
Let $\mathcal{H}$ be the Hilbert space representing one such spin, we denote by $\LL(\mathcal{H})$ the space of linear operators on $\mathcal{H}$.
Let the   chain initially be in an arbitrary state $\rho_0 \in \LL\left(\mathcal{H}^{\otimes N}\right)$ and evolve in time according to a unitary time evolution generated by a finite range interaction Hamiltonian $H=\sum_i h_{i}$, where the sum runs over  all spins  $i$ in the chain,  each term $h_i$ has a finite range, $R$, and acts on at most $n(R)$  spins contained in a ball of radius $R$ around spin $i$. 
We assume that the interaction strength is bounded by  $\norm{h}:=\sup_i \norm{h_{i}}$ for any size of the lattice.
 For brevity we present the proof in the case of strictly finite range Hamiltonians. The same proof works for Hamiltonians with sufficiently strongly decaying interactions, precisely, for quasi-local Hamiltonians as per Ref.~\cite{Marien2016}.

At times $(t_1,t_2,\ldots)$ a quantum instrument   acts separately on each spin (for now,  assume that this happens instantaneously, we shall relax this assumption in  \cref{thm:cont_meas}).
The different  measurements (we use the terms measurement, quantum operation and quantum instrument interchangeably) are   performed at spacetime points $(x,t)$, where $x$ is the position of the spin in the chain and $t$ is the time of the measurement.
We purify  each measurement device, thereby  associating to each spacetime point an ancillary quantum system. 
The state of the ancillas after interacting with the spins is given by the   tensor network state shown in \cref{fig:TN}. 

Let $\mathbf{A}$ be a spacetime region comprised of $X$ neighboring  spins and spanning a time interval of duration $\Ttot=\Tsteps\Dt$, where $\Tsteps$ is the number of time steps and $\Dt$ is the  length of time evolution between measurements (see \cref{fig:TN}).
We measure the spatial extent of a system in units of the lattice spacing, so that in one dimension, length  is equal to the number of spins. For ease of notation, the time intervals between measurements are taken to be equal.  The same result holds for arbitrary time intervals. 
Alice controls the measurements inside the region $\mathbf{A}$ and Bob the ones outside of it.

We shall formulate  our bound in terms of  the quantum mutual information between the ancillas associated to  the measurements performed  by Alice   	and the rest of the system (which includes Bob's ancillas  and the physical spins at the end of the measurement process).

The quantum mutual information   of a bipartite quantum  system in a state $\rho \in \LL(\mathcal{H}_A\otimes\mathcal{H}_B)$ is given by~\cite{nielsen_chuang_2010}:
\begin{equation*}  
I(A:B)_\rho = S(\rho_A)+S(\rho_B) - S(\rho ) \ ,
\end{equation*}
where $S(\rho) = -\tr \rho \log \rho$ is the von Neumann entropy and $\rho_A=\tr_B\rho $ is the reduced state of the system $A$.

 Before stating our results, we demonstrate that the    quantum mutual information between the ancillas purifying the agents' instruments is an upper bound on the operational quantities of interest, namely: 
(a)
the classical mutual information between measurement outcomes of Alice and Bob;  
and 
(b)
the classical channel capacity from Alice's instrument settings to Bob's outcomes when signaling  is considered. 
Further note that  the distillable entanglement is also bounded by the quantum mutual information~\cite{Devetak207}.

Quantum mutual information is non-increasing  with respect to applying  completely positive and trace preserving  (CPTP) maps  separately  on each subsystem~\cite{nielsen_chuang_2010}.
In particular, tracing out parts of subsystems does not increase mutual information. 
We therefore  assume w.l.o.g. that the initial state of the spins is pure. If it were a mixed state, we could consider a purification, and the mutual information would only be higher. For a pure system the quantum mutual information is equal to twice the von Neumann entropy of the reduced state. 

The same monotonicity property further  implies  that the  \textit{classical} mutual information between  the probability distributions for the     outcomes of measurements performed  by  Alice and those performed     by Bob,  is bounded by  the quantum mutual information (point a). 
To see this recall that those   measurement outcomes are obtained by performing projective measurements on the ancilla systems  
and   note that the  map which transforms the state of the ancillas to a diagonal density matrix which encodes the probabilities for the  measurement's outcomes  is CPTP.

When considering a signaling scenario in which  Alice can choose an instrument setting $a$ with probability $p(a)$, we can encode this probability distribution in the state of an additional ancilla $\ket{p}=\sum_a \sqrt{p(a)}\ket{a}$ and apply different unitaries conditioned on this state.
As   above, we can then bound the classical  mutual information between $p(a)$ and the probability distribution governing  Bob's outcomes. 
 If our bound does not depend on the distribution  $p(a)$ (as will be the case in our \cref{thm:main,thm:cont_meas}, since both do not depend on  the initial state of the  ancillas), then the classical channel capacity~\cite{nielsen_chuang_2010} from Alice's choice of settings to Bob's outcomes is   also bounded (point b above).

 Note that in the case of one spatial dimension, if  the evolution between time steps is given by a matrix product operator~\cite{CIRAC2017} (this is exactly the case when the spins' evolution is governed by aquantum cellular automaton~\cite{Schumacher2004}, since those have been shown to   coincide with  translationally invariant matrix product unitary operators~\cite{Cirac2017a}),  
then the proof of a spacetime area law bound on the  quantum mutual information follows immediately from the representation of the state of the ancillas as a two-dimensional tensor network state in \cref{fig:TN}. To see this note  that tensor network states obey an area law bound on the entanglement entropy of subsystems~\cite{Verstraete2008}. Replacing the time evolution operators in \cref{fig:TN} by matrix product operators we see that the number of bonds cut by the red line defining Alice's subsystem equals the area of the spacetime region $\mathbf{A}$ ($=2\Ttot+2X$). Thus we obtain  an  upper  bound on the entropy of Alice's reduced density matrix which is proportional to $|\partial \mathbf{A}|$.

We shall now state our     results   for spin lattices of any spatial dimensions. 

\begin{result} \label{thm:main}
Let a   spin  lattice in $D$ dimensions, with each spin described by a $d$-dimensional Hilbert space, initially be  in a state $\rho_0$, and let the spins   evolve in time   according  to a finite range    Hamiltonian $H = \sum h_i$ with range $R$ and bounded interaction strength  $\norm{h}=\sup_i\norm{h_i}$ independently of system size.
 Let arbitrary   quantum instruments be applied individually on each spin at times $(t_1,t_2,\ldots)$.
Let  $\Sigma$ be a   subset of the  lattice and let $(t_\alpha,t_\beta)$ be a time interval so that together they define a  spacetime region $\mathbf{A}=\Sigma\times (t_\alpha,t_\beta)$. Let $\rho$ be the state of the combined system of spins and ancillas at the end of the measurement process;
 then there exists a constant $C>0$ which depends only on  $D$ and  $R$, such that the following bound holds for the quantum mutual information between the ancillas (denoted $A$) corresponding to measurements performed inside the region  $\mathbf{A}$ and the rest of the system $\bar{ A}$:
 \begin{equation*}  
 	I( {A}:\bar{ {A}})_\rho \leq C \norm{h} |\partial \mathbf{A}| \log d  \ ,
 \end{equation*}
where      $|\partial \mathbf{A}| = 2 |\Sigma|  + \Ttot |\partial \Sigma|$, with $\Ttot= t_\beta-t_\alpha $ (assuming equal time steps $\Ttot=\Tsteps \Dt$) and where $\partial\Sigma$ denotes the boundary of the region $\Sigma$ and $|\cdot|$ counts the number of elements in a set. 
\end{result}

 The proof of \cref{thm:main} is given in the methods section below. 
 From the proof  it will become clear that the same bound holds even when  both Alice and Bob are allowed to perform \textit{collective}   measurements within their regions at each time step, as well as, when they    are allowed to reuse  ancillas from previous time steps for their measurements. This observation leads to the   following extension of   \cref{thm:main} to a setting where  measurements are continuous in time,  particular cases of which are entanglement harvesting scenarios. 
	
\begin{result} \label{thm:cont_meas}
	Let $a$ and $b$  be arbitrary finite dimensional  ancillary systems; let $\Sigma$ be a subset of the spin lattice and  $\bar{\Sigma}$ be its complement; and let $H_{0}$ be a finite range Hamiltonian  for the spins with range $R$ as in \cref{thm:main}. 
	Let the system initially be in a pure state $\ket{\psi}_{\Sigma \bar{\Sigma}} \otimes\ket{0}_a\otimes\ket{0}_b$.
	Let $ \TT=(t_\alpha,t_\beta)$ be a time interval of length $T=t_\beta-t_\alpha$ and let the time dependent Hamiltonian of the system be 
	\begin{equation*}
	H= H_{0} + H_{b,\bar{\Sigma}} +(1-\XX_{\TT}) H_{b,\Sigma} + \XX_{\TT} H_{a,\Sigma}  \  , 
	\end{equation*}
	where $H_{X,Y}$ denotes a time dependent  interaction Hamiltonian between systems $X$ and $Y$; and $\XX_{\TT}(t)$ the indicator function of the interval $\TT$ (equals unity for $t\in \TT$ and  zero otherwise).
	For any $t\geq t_\beta$,  the quantum mutual information between the ancillas  $a$ and $b$ satisfies:
	\begin{equation*}  
	I(a:b)(t)\leq C \norm{h}(2|\Sigma|+T|\partial \Sigma|)  \log d  \  ,
	\end{equation*} 
were the constant $C$ depends only on  $D$ and  $R$ as in \cref{thm:main};
\end{result}

 The proof of \cref{thm:cont_meas} is given in the methods section below.

 \section*{Discussion}
 We have considered local operations performed on a lattice spin system   evolving under local dynamics. 
 We  showed that the mutual information between  outcomes of local measurements is  bounded by a  {spacetime area	 law}. In particular, the amount of classical information that an agent, localized within a spacetime region, can send outside or infer about the outside  is at most proportional to the  area of the region's boundary. 
 Agents trying to harvest entanglement from the spin lattice by coupling detectors to it will run into the same bound.

 The results obtained in the present article complement the results in Ref.~\cite{Bravyi2006}. There,  the Lieb-Robinson bound is used to determine \textit{where}   information  can travel in (discretized) spacetime. Our result bounds \textit{how much} of it can be shared between spacetime regions.
 Furthermore, this work is a  rigorous demonstration  of an intuitively compelling idea that provides a link between quantum information and spacetime geometry. This idea, that information travels across boundaries of spacetime regions, is at the heart of a recent  approach     to the foundations of quantum theory \cite{Oeckl2016,OECKL2003318} where it is used to motivate the association of   quantum states to  boundaries of arbitrary spacetime regions.

 The proven  bounds  hold  independently of the instruments used, the details of the purification and of the dimensions of the ancillary systems, and can be, therefore, regarded   as an intrinsic  property of the dynamics.  Based on this observation, we suggest that the maximal mutual information between ancillary systems of measurement devices can be used as a measure of correlation intrinsic to a general quantum process. 
We discuss this further  in Appendix~\ref{app:process_matrix}. There we define the proposed measure precisely within the process matrix formalism \cite{Oreshkov2012}. This discussion is intended mainly 
	for readers interested in 
	various operational approaches to multipartite signaling quantum correlations{~\cite{Chiribella2008, Chiribella2009, Hardy2012, Cotler2018,Modi2018,Oreshkov2012}}.

 We    have  restricted our analysis to finite-dimensional systems in a non-relativistic setting. 
	In relativistic quantum field theory, the dimension of the local Hilbert space at one spacetime point is infinite, which makes the results presented here unapplicable a priori.  However, it seems reasonable to expect   that a similar result should hold for relativistic quantum field theories which can be accurately simulated by a spin lattice with local interactions (see e.g.\ \cite{Jordan2012,Martinez2016,Preskill:2018fag}). We expect that by   restricting   the    admissible initial states and Hamiltonians, similar bounds could be obtained for regularized quantum field theories. We leave these questions for future study.

\section*{Methods} 
\begin{figure}[h]
 		\begin{tikzpicture}[baseline=-1mm, scale=1, every node/.style={transform shape}]
 		\draw  (0.3,0) arc (120:240:1);
 		\draw (0.3,-1.73)--(0.3,0)  ;
 		
 		\draw (0.3,-.5)--(6.5,-.5);  \draw (7.5,-.5)--(11.5,-.5); 
 		\node at (7,-.5) {\large{$\ldots$}};
 		\draw (0.3,-1.23)--(6.5,-1.23);  \draw (7.5,-1.23)--(11.5,-1.23); 
 		\node at (7,-1.23) {\large{$\ldots$}};
 		\draw (1.1,-3)--(6.5,-3);  \draw (7.5,-3)--(11.5,-3); 
 		\node at (7,-3) {\large{$\ldots$}};
 		
 		\node at (11.8,-3) {\large{$\ldots$}};
 		\node at (11.8,-1.23) {\large{$\ldots$}};
 		\node at (11.8,-.5) {\large{$\ldots$}};
 		
 		\node at (.3,-3) {\large{ $\Ket{0}^{\otimes |A|}$}};

 		\foreach \i in {2,4,6,8,10 }
 		\pic[pic text = {\large$U$}]  at (\i,-.87) {greygate};
 		\foreach \i in {1,3, 11 }
 		\pic[pic text =] at (\i,-.87) {whitegate};
 		\foreach \i in {5,9 }
 		\pic[pic text =] at (\i,-2.12) {whitegateLONG};
 		\foreach \i in {5,9 }
 		\pic[pic text =] at (\i,-.5) {whitegateONE};

 		\node  at (-.75,-3) {\Large{$A$}};
 		\node  at (-.75,-.5) {\Large{$C$}};
 		\node  at (-.75,-1.23) {\Large{$B$}};
 		\node  at (1,-3.75) {\large{Time step:}};
 		\node  at (3.4,-3.75)   {\large{$t_0$}};
 		\node   at (6.4,-3.75)   {\large{$t_1=t_0+\Dt$}};
 		\node   at (10.4,-3.75) {\large{$t_\Tsteps = t_0+\Tsteps\Dt$}};
 		\draw [style=thin,style=dotted] (3.5,-3.5)--(3.5,0);
 		\draw [style=thin,style=dotted] (5.5,-3.5)--(5.5,0);
 		\draw [style=thin,style=dotted] (9.5,-3.5)--(9.5,0);

 		\end{tikzpicture}
 	\caption{The circuit representation of \cref{fig:TN}. As described in the main text, subsystem $A$ consists of the ancillas inside the region $\mathbf{A}$ (encircled by the dashed red line in \cref{fig:TN});
 		subsystem $B$ consists of the  spins  which Alice measures;
 		subsystem $C$  consists of the rest of the spins {and} ancillas. The white gates correspond to the action of the instruments (unitary interaction between spins and ancillas). The shaded gates represent the unitary time evolution. The time axis is aligned with that of \cref{fig:TN}. 
 	} 
 	\label{fig:circ}
 \end{figure}
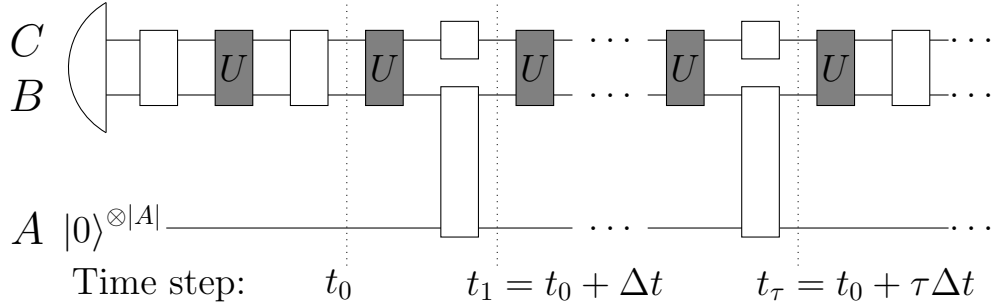

We first prove \cref{thm:main}. 
 As in the results section, we begin by considering the one-dimensional case. 
Consider the following division of  the spins and ancillas into  three sets:
 ($A$) the ancillas inside the region $\mathbf{A}$ (encircled by the dashed red line in \cref{fig:TN});
	($B$)  the  spins     which Alice measures;
 and	($C$)  the rest of the spins \textit{and} ancillas.
 The sequence of measurements can now be represented by the   circuit diagram  in \cref{fig:circ}, which is   key to understanding how the subsystem $A$ gets entangled with the rest of the system.

Denote by $t_0$ the time of the last measurement which precedes the measurements inside $\mathbf{A}$, let $t_m:=t_0 +m\Dt$ and let $t_{\mathrm{f}}$ be the time of the final measurement.  From   $t_\Tsteps$ onward the system $A$ does not interact with the system $BC$ (see \cref{fig:circ}), therefore the entropy of $A$ at the end of the entire process, $S_A(t_{\mathrm{f}})$, which is the quantity which  we wish to bound, is equal to $S_A(t_\Tsteps)$. 

In the time interval $(t_0,t_\Tsteps)$ systems $AB$ and $C$ interact only via the time evolution operators   acting on the physical spins. The Small Incremental Entangling (SIE) theorem proved in   Refs.~\cite{VanAcoleyen2013,Marien2016} bounds the rate of entanglement generation by  the time evolution  operator and allows us to bound the increase of $S_C=S_{AB}$ with each time step.  
	 Using a Trotter expansion of the time evolution operator, it is easily shown 	 that  the only terms in the Hamiltonian  able to generate entanglement (increase $S_C$) are the ones that act across the boundary between systems  $B$ and $C$. When there are $M$ such terms, i.e.\ when the Hamiltonian decomposes as $H= H_C+H_B+\sum_{i=1}^M H^i_{CB}$, and when each term is supported on at most $n$ spins, the SIE theorem implies that the change of the entropy $S_C$ after evolving for a (finite) period of time $\Dt$ is bounded by
	\begin{equation*}  
	 \Delta S :=  | S_C(\Dt) - S_C(0) | \leq c \Dt   M (n-1) \norm{h} \log d \ ,
 \end{equation*}
 	 where $c$ is a numerical constant from  the SIE theorem.
 
In the $1D$ case, $M$ - the number of interaction terms  between systems $B$ and $C$,   is equal to $2(n-1)$.  
Next, bound the total increase in  $S_C$ in the time interval of interest by

		\begin{equation} \label{eq:timeStepBound}
	\begin{split}
	S_{AB}(t_\Tsteps) &=S_{C}(t_\Tsteps) \leq 
				S_{C}(t_{\Tsteps-1})+\Delta S   \\
				&\leq S_{C}(t_{\Tsteps-2})+2\Delta S \leq \ldots \leq S_{C}(t_0)+\Tsteps\Delta S \\
					&= S_{AB}(t_0)+\Tsteps\Delta S  = S_B(t_0) +\Tsteps\Delta S \ ,
	\end{split}
	\end{equation} 
where in the last step we used the fact that at time $t_0$ the system $AB$ is in a product state and the state of $A$ is pure. Using the triangle inequality for $S_{AB}$~\cite{araki1970} and \cref{eq:timeStepBound} we obtain
		\begin{equation*}
| S_A(t_\Tsteps) - S_B(t_\Tsteps) | \leq S_{AB}(t_\Tsteps) \leq S_B(t_0) +\Tsteps\Delta S \ . 
\end{equation*} 
Recalling  that $S_A(t_\Tsteps )=S_A(t_{\mathrm{f}})$ we obtain
		\begin{equation*}
S_A(t_{\mathrm{f}}) \leq  S_B(t_0)  +  S_B(t_\Tsteps) +\Tsteps\Delta S \leq 2 \log (d_B) +\Tsteps\Delta S  \ ,
\end{equation*} 
where $d_B$ is the dimension of $\mathcal{H}_B$, i.e.\ $d_B=d^X$, and we bounded $S_B(t)$ by its maximum possible value.
  Plugging in the bound $\Delta S$ we obtain the desired area law
		\begin{equation*}
\begin{split}
S_A(t_{\mathrm{f}}) &\leq
 2  \log (d^X) +2c\Ttot (n-1)^2 \norm{h} \log (d )      \\
&\leq C(n)\norm{h} ( 2X + 2T) \log (d )
\ ,
\end{split}
\end{equation*}
where $C(n ) := 2c (n-1)^2  $ and $c$ from the SIE theorem ({w.l.o.g.} $ \norm{h}c>1$). This proves   \cref{thm:main} in the case of one spatial dimension (since $|\partial \mathbf{A}| =2X + 2T$).

The proof is essentially the same  {when space is $D$-dimensional}. The circuit diagram representation in \cref{fig:circ}  holds true. It remains only to compute $M$, the number of interaction terms across the boundary between systems $B$ and $C$.   Let  {$n(D,R)$} be the number of spins inside a ball of radius $R$.  For a finite range Hamiltonian with range $R$, the number of interaction terms acting on a single spin in the lattice is then at most   {$n(D,R)$}. Ignoring multiple   {counting} of the same  terms, we can bound $M$ by   {$|\partial \Sigma| \cdot n(D,R)$}. Plugging this into the above $\Delta S$ and using $S_B \leq |\Sigma| \log d$ in \cref{eq:timeStepBound} proves   \cref{thm:main}.

For the proof of \cref{thm:cont_meas} first note  that \cref{fig:circ} does not change neither if   both Alice and Bob are allowed to perform \textit{collective}   measurements  at each time step, nor if they   reuse  ancillas from previous time steps.
Further note that 
only the time interval $\TT$ is  of interest because before and after it the ancilla $a$ does not interact with any other system.
 Next we use the Trotter decomposition  in order to arrive to a setting with discrete time steps, in which we can apply the same reasoning as in the proof of \cref{thm:main}.
 Split  the  time interval $\TT$ into $m$ equal intervals,   and  in each one approximate  the time evolution operator using the   (first order)  Lie-Trotter-Suzuki product formula~\cite{Wiebe2010}. The approximate time evolution operator takes the form:
 	\begin{equation*}
 \begin{split}
 \prod_{l=1}^{m}  
 e^{-i \frac{T}{2m} H_{0}(t_{l })} 
 e^{-i \frac{T}{2m} H_{b,\bar{\Sigma}}(t_{l })}  
 e^{-i \frac{T}{2m}  H_{a,\Sigma} (t_{l }) }    \times
 \\
 e^{-i \frac{T}{2m}  H_{a,\Sigma} (t_{l }) }    
 e^{-i \frac{T}{2m} H_{b,\bar{\Sigma}}(t_{l })}  
 e^{-i \frac{T}{2m} H_{0}(t_{l })} 
 \ ,
 \end{split}
 \end{equation*}
 where $t_l$ is the middle of the $l$-th time interval $t_l = t_\alpha +\frac{l\times T}{m} - \frac{T}{2m}$.
 When applied to the initial state of the system, this sequence  of unitary operators is described by the same circuit diagram \cref{fig:circ} with subsystems: $A:=a$, $B:=\Sigma$ and $C:=\bar{\Sigma}\cup b$. I.e.\ the operators 
 $e^{-i \frac{T}{2m} H_{0}} $ are  represented in \cref{fig:circ} by  the gray gates; and  $e^{-i \frac{T}{2m} H_{b,\bar{\Sigma}}}$  and $ e^{-i \frac{T}{2m}  H_{a,\Sigma} } $ by the white gates. 
 For any $m$, \cref{fig:circ} and the entropy inequalities in the proof of \cref{thm:main}, imply that the von Neumann entropy of the ancilla $a$ is bounded by the spacetime area law corresponding to  the box $\Sigma \times \TT$   (there are $\Tsteps=2m$ time steps of duration $\frac{T}{2m}$).   
 This bound holds true for any $m$ and does not depend on its value. In Ref.~\cite{Wiebe2010} it was shown that for a sufficiently smooth time dependent Hamiltonian, the  error of the Lie-Trotter-Suzuki formula vanishes as $m$ tends to infinity. Using the  continuity of the von Neumann entropy with respect to the state, we conclude that the same bound  holds   for the exact time evolution. The bound on the quantum mutual information follows from the fact that for pure states it is equal to twice the von Neumann entropy of the reduced state, and the fact that it is non-increasing with respect to tracing out subsystems~\cite{nielsen_chuang_2010}.

\section*{Acknowledgments}
We thank Frank Verstraete for bringing the SIE theorem to our attention and Marius Krumm, Alessio Belenchia, Simon Milz and Jordan Cotler for fruitful discussions.   {We acknowledge the support of the Austrian Science Fund (FWF) through the Doctoral Program CoQuS and the project I-2526-N27 and the research platform TURIS, as well as support from the European Commission via Testing the Large-Scale Limit of Quantum Mechanics (TEQ) (No. 766900) project. P.A.G. acknowledges support from the Fonds de Recherche du Qu\'{e}bec -- Nature et Technologies (FRQNT). This publication was made possible through the support of a grant from the John Templeton Foundation. The opinions expressed in this publication are those of the authors and do not necessarily reflect the views of the John Templeton Foundation. } 
This is a pre-print of an article published in npj Quantum Information. The final authenticated version is available online at: 
https://doi.org/10.1038/s41534-019-0171-x.

  \pagebreak
 
 \begin{appendices}

 \section{Maximal Bipartite Correlation }
 	\label{app:process_matrix}
 	 (This appendix appears online as a supplemental discussion to the article, and is available also at https://doi.org/10.1038/s41534-019-0171-x.)
 	 
 We  elaborate on our proposal, mentioned  in the main text of the article, to define a maximal measure of bipartite correlation for general quantum processes. 
 The following discussion is presented in the framework of the process matrix formalism~\cite{Oreshkov2012}. It is, however, applicable in many other operational approaches for dealing with multipartite signaling quantum correlations{~\cite{Chiribella2008, Chiribella2009, Hardy2012, Cotler2018,Modi2018}} since these are  closely related  to the process matrix formalism.
 
 The operational treatment of multi-time correlations in quantum theory  dates back to Lindblad~\cite{lindblad1979}  and Accardi et al.~\cite{Accardi1982}; there, a quantum stochastic process is described by a multi-time correlation matrix which encodes, for a given choice of measurements, the probabilities for all the possible outcomes at different times. 
 Recent developments of this approach~\cite{Chiribella2008, Chiribella2009, Hardy2012, Modi2018,Oreshkov2012}   describe the most general process, involving several agents,   as a functional acting on the space of all possible quantum operations available to them. Such processes can treat   combined spatio-temporal scenarios and can thus  be seen as describing a spacetime in which the agents are embedded, or more generally,  a causal structure (see~\cite{Oreshkov2012, Cotler2018, Fitzsimons2015, Horsman20170395,Hardy2005,Hardy2007,Hardy2009}).

 The process matrix formalism allows to compute the joint probabilities for the outcomes of quantum experiments performed in local laboratories, without needing to pre-assume a definite causal order between the different laboratories. It is general enough to describe any causally ordered scenario~\cite{Chiribella2009, Costa2016}, as well as experimentally relevant non-causal processes~\cite{Chiribella2013, Araujo2015, OpticalQuantumSwitchProcopio, OpticalQuantumSwitchRubino, Goswami2018, Oreshkov2018, Oreshkov2016}; it also allows for the possibility of the violation of so-called causal {inequalities}~\cite{Oreshkov2012,Branciard:2016aa, Abbott:2016aa,Miklin_ineq:2017, Oreshkov2016}, a counterintuitive and so far unobserved phenomenon.

 We briefly review the basics of the process matrix formalism. We refer the  reader to the original reference~\cite{Oreshkov2012}, as well as Refs.~\cite{Araujo2015, Araujo2017} for more complete introductions. Our choice of convention for the Choi-Jamio\l kowski isomorphism~\cite{Jamiolkovski, Choi1975} follows that of Ref.~\cite{Araujo2017}.
 
 Let $\HH$ be a finite-dimensional Hilbert space, we denote by $\BB(\HH)$ the space of linear operators acting on $\HH$. We use $\HH_{AB}$ to denote the tensor product of  two Hilbert spaces $\HH_A \otimes \HH_B$. 
 We shall use superscripts  such as    $A_I,A_O,B_I, B_O$ to indicate the Hilbert spaces on which operators act. 
 
 To every linear map $\mathcal{M}: \BB(\mathcal{H}_{A_I}) \to \BB( \mathcal{H}_{A_O})$ corresponds a \textit{Choi matrix} $M \in \BB(\mathcal{H}_{A_I A_O})$ which is given by
 \begin{equation*}
 M^{A_I A_O} = \sum_{i j = 1}^{\mathrm{dim} \HH_{A_I}} |i \rangle \langle j|^{A_I} \otimes \mathcal{M} (|i \rangle \langle j|)^{A_O} \ .
 \end{equation*}
 The inverse of this isomorphism is given by the formula
 \begin{equation*}
 \mathcal{M}(\rho) = \tr_{A_I} \left(M^{A_I A_O} \cdot (\rho^{A_I})^{T} \otimes \II^{A_O} \right),
 \end{equation*}
 where $T$ denotes the transposition in the computational basis. The above equation can be used to show that $\mathcal{M}$ is trace-preserving iff $\tr_{A_O} M^{A_I A_O} = \II^{A_I}$. Choi's theorem~\cite{Choi1975} states that $M^{A_I A_O}$ corresponds to a completely-positive (CP) map if and only if $M \geq 0$.

 A quantum instrument is defined as a collection $\{ \MM_{a} \}$ of CP maps,  labeled by a finite index  $a$,  which sum to a CPTP map. For an initial state $\rho$,  outcome $a$ is recorded by the instrument with probability $\tr{\MM_a (\rho)}$ and the post measurement state is given by $\MM_{a}(\rho) /  \tr{\MM_a (\rho)}$.
 
 We shall define process matrices in the case of two parties, whose local laboratories  have input Hilbert spaces $\HH_{A_I}, \HH_{B_I}$, and output Hilbert spaces $\HH_{A_O}, \HH_{B_O}$ respectively. 
 A process is a positive linear functional $ \WW$ acting on instruments to produce probabilities for the measurement outcomes. For a given  choice of instruments for $A$ and     $B$, we can use the  Choi matrix of the process,  $W \in \BB( \HH_{A_I  A_O B_I B_O})$, to    represent the probability for  obtaining outcomes $a$ and $b$     by 
 \begin{equation*}
 p(a,b) =  \ \tr \left[ \Big(M^{A_{I} A_O}_{a}\otimes M^{B_I B_O}_{b} \Big) W^T \right] \ .
 \end{equation*}
 
 The requirement that a process matrix  $W$  should produce valid probabilities  for all choices of local quantum instruments (also in the case when the parties share a pre-established entangled state) forces it to satisfy $W\geq 0$, $\tr W =d_{A_O} d_{B_O} $, and in addition, it restricts the process to lie in a subspace of  $  \BB( \HH_{A_I  A_O B_I B_O})$~\cite{Araujo2015, Araujo2017}.
 The generalization of the definition to more than two parties is straightforward.

 \textit{A general measure of correlations for quantum processes} - 
 Let $W$ be an $N$-partite process matrix, with parties labeled from $1$ to $N$. To each party $j$,   associate  an input Hilbert space $\HH_{I_j}$ and  an output Hilbert space $\HH_{O_j}$. In addition, to each party associate an ancillary system of arbitrary dimension, in an initial state $\ket{0}\in\HH_{ I_j^\prime}$. Let each of the parties  probe the process by having his ancilla  interact unitarily with the process, i.e.\ apply a joint unitary $U_j:\HH_{I_j}\otimes \HH_{I^\prime_j} \rightarrow \HH_{O_j}\otimes \HH_{O^\prime_j} $. 
 
 
 Given a bipartition of the parties into two sets, after each party interacted with the process,  we can consider the mutual information of the state of the ancillas with respect to the bipartition, and   to optimize over the probing schemes as to obtain the maximum mutual information possible.
 Due to the monotonicity property of mutual information with respect to the  partial trace operation, the same maximum is  obtained if, instead of the unitary probing scheme, we allow each party to apply a local CPTP map $\MM_j: \LL(\HH_{I_j}) \to \LL(\HH_{O_j} \otimes \mathcal{H}_{O_j}^\prime)$ and optimize over all such maps. We shall, therefore, use this description.
 
\begin{figure}[h]
 	\begin{tikzpicture}[baseline=-1mm,scale=.65, every node/.style={transform shape}]
 	
 	\draw (1,1)--(2.5,1)--(2.5,2)--(-2.5,2)--(-2.5,1)--(-1,1)--(-1,-1)--(-2.5,-1)--(-2.5,-2)--(2.5,-2)--(2.5,-1)--(1,-1)--(1,1);\
 	\draw [-latex] (2,-1)--(2,-.5);
 	\draw [-latex] (-2,-1)--(-2,-.5);	
 	
 	\draw [-latex] (2,.5)--(2,1);
 	\draw [-latex] (-2,.5)--(-2,1);
 	
 	\draw [-latex] (3,.5)--(3,2.5);
 	\draw [-latex] (-3,.5)--(-3,2.5);
 	\pic at (2.5,0) {box};
 	\pic at (-2.5,0) {box};
 	
 	\node at (0,0) {\Large{$W$}};
 	\node at (-2.5,0) {\Large{$M_A$}};
 	\node at (2.5,0) {\Large{$M_B$}};
 	
 	\node at (-1.7,.75) {\large{$A_O$}};
 	\node at (2.3,.75) {\large{$B_O$}};
 	
 	\node at (-1.7,-.75) {\large{$A_I$}};
 	\node at (2.3,-.75) {\large{$B_I$}};
 	
 	\node at (-2.7,2.2) {\large{$A^\prime_O$}};
 	\node at (3.3,2.2) {\large{$B^\prime_O$}};
 	\end{tikzpicture}
 	\caption{Graphical representation of~\cref{eq:resulting_G}, in the bipartite case, for calculating the final state of the ancillas after all the parties interacted with the process. Here $M_A$ and $M_B$ are the Choi matrices of the local CPTP maps $\mathcal{M}_A$ and $ \mathcal{M}_B$.}
 	\label{fig:process}
 \end{figure}
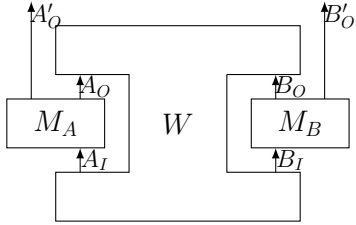

 For a two party process  matrix, the  state of the ancillas at the end of the experiment, $\rho_{\MM}$, is given by the following formula,  whose graphical representation is Figure~\ref{fig:process}.
 
 \begin{equation}
 \label{eq:resulting_G}
 \begin{split}
 &\rho_{\{\MM\}}=\\ 
 &\tr_{A_I A_O B_I B_O} \left(W^{T} (M_A^{A_I  A_O A_O^\prime} \otimes M_B^{B_I B_O B_O^\prime}) \right)  \ ,
 \end{split}
 \end{equation}
 where $M_A$ and $M_B$ are the Choi matrices of the local CPTP maps applied by $A$ and $B$ respectively, and the subscript ${\{\MM\}}$ emphasizes the  dependence of the state  on the choice of such maps.

 Let  $(Z : \bar{Z})$ be  a bipartition of the $N$ parties into two sets $Z$ and $ \bar{Z}$.
 By maximizing the mutual information of the final state of the ancillas with respect to the bipartition $I_{\rho_{\MM}}(Z : \bar{Z})$, over all probing schemes $\MM$, we get a bound on the maximal amount of correlations that the process allows across this bipartition.
 This leads us to define the following intrinsic measure of the strength of correlations allowed by a process:
 
 \begin{equation*}
 \mathcal{C}_W(Z : \bar{Z}) := \sup_{\MM} I_{\rho_{\MM}}(Z:\bar{Z}) \ ,
 \end{equation*}
 where the supremum  is taken over all CPTP maps $\{ \MM_j \}_{j=1}^{N}$ and all dimensions for the ancillary output Hilbert spaces.	
 
 Notwithstanding the fact that the ancillary dimensions can increase indefinitely, this quantity is finite.
 It  is intuitive that only a finite amount of information can be shared by the two parties, as their ancillas are initially in a product state and  the correlations between  them are created by the process $W$, which has finite dimensions.
 It can be shown that the   rank of the final ancilla state $\rho_{\{\MM\}}$ in  \cref{eq:resulting_G} is bounded by the products of the dimensions of $\HH_{A_I},\HH_{A_O},\HH_{B_I}$ and $\HH_{B_O}$.
 Consider a purification of  of the   local CPTP maps $\MM_A$ and $\MM_B$ (there is no loss of generality since tracing out the purifying system will only decrease the mutual information), and apply a Schmidt decomposition to the vectors defining   their (pure) Choi matrices.
 \begin{equation*}
 \begin{split}
 &M_A^{A_IA_OA_O^\prime} = \\
 &\sum_{i }^r 
 \ket{\phi_i}^{A_IA_O} \otimes \ket{\psi_i}^{A_O^\prime}
 \sum_{i }^r
 \bra{\phi_j}^{A_IA_O} \otimes \bra{\psi_j}^{A_O^\prime} \ , 
 \end{split}
 \end{equation*}
 where 
 the Schmidt rank $r$ is not greater that $\min(d_{A_I}\times d_{A_O},d_{A_O^\prime} )$.
 Plugging this into \cref{eq:resulting_G} gives  the desired bound.

 The quantity $\CC_W(Z:\bar{Z})$ is exactly the quantity for which a  spacetime area law  bound was obtained in the main article,  in the case when $W$ describes a local Hamiltonian evolution.

 It is insightful to   consider  some examples to illustrate the meaning of $\CC_W$ for simple processes.
 In the case of two parties, which we label by $A$ and $B$, let $W = \omega^{A_I B_I} \otimes \II^{A_O B_O}$ be a state shared between the two parties. In this case it is straightforward to see that $\CC_W(A:B)$ is just the usual quantum mutual information of the state $\omega$.
 Another important example is when the process is a channel from $A$ to $B$. In this case the process matrix  is given by  $W^{A_O B_I}$, the Choi state of the channel $\WW: \LL(\HH_{A_O}) \to \LL(\HH_{B_I})$. Here $A_I$ and $B_O$ are assumed to be the trivial Hilbert spaces and, therefore, the CPTP map on $B$'s side, $\MM_B^{B_I \to B^\prime_O }$, can be disregarded as it can only decrease the mutual information.  $\mathcal{C}_W(A:B)$  is then given by 
 the supremum over all states $\omega\in \BB(\HH_{A_O A^\prime_{O}})$, of the quantum mutual information of the state  
 $ (\II^{A_O^\prime} \otimes \WW^{A_O\to B_I}) \omega^{A_O A^\prime_{O}} $, i.e.\ the state obtained by sending part of $\omega$ through the channel.
 This quantity is known to be equal to the entanglement-assisted classical capacity of the channel $\mathcal{W}$~\cite{Bennett2002}, which is evidently an upper bound to the ordinary classical capacity of $\mathcal{W}$.
 
 One interesting probing scheme, which is considered in Refs.~\cite{Cotler2018,Araujo2017}, consists of all parties performing 
 $\MM_A (\rho) = 
 \rho^{A_{O}^{\prime 1}} \otimes \ket{\Phi^+} \bra{\Phi^+}^{A_O A_{O}^{\prime 2}}$, where $\ket{\Phi^+}$ is the maximally entangled state and $A'_{O} := A_{O}^{\prime 1} \otimes A_{O}^{\prime 2}  \cong A_I \otimes A_O$. 
 In this case, the final state of the ancillas is equal (up to normalization), to the process matrix $W$.
 This shows that $\CC_W$ is at least as large as the mutual information of the state $\rho = W / \tr(W)$. However, one can find examples showing that $\CC_W$ is in general strictly larger than that. Indeed, let $d_{A_I} = d_{A_O} = d_{B_I} = 2$, $d_{B_O} = 1$, and take the (causally-ordered) process
 \begin{equation*}
 \label{eq:W_counter_ex}
 W =|\Phi^+ \rangle \langle \Phi^+|^{A_I B_I} \otimes |0 \rangle \langle 0|^{A_O} + \frac{1}{4} \II^{A_I B_I} \otimes |1 \rangle \langle 1|^{A_O}.
 \end{equation*}
 The mutual information of $W/\tr(W)$, across the  bipartition $(A:B)$ is 1 bit, while for the probing scheme $\MM_A(\rho) = \ket{0}\bra{0}^{A_O} \otimes \rho^{A_{O}^\prime}$, $\MM_B(\rho) = \rho^{B_{O}^\prime}$, the final state of the ancillas has a mutual information of 2 bits.
 
 In conclusion, we comment that 
 it has recently been suggested that spatial geometry can be reconstructed from the entanglement structure
 of certain quantum states~\cite{Carroll2017, Kempf2018, Jackobson2016, VanRaamsdonk2010}.   
 Here we proposed a measure of maximal bipartite correlation which is intrinsic to general quantum processes.
 We propose that based on this measure of correlation, similar methods could be used to reconstruct the spacetime geometry  structure of some class of quantum processes.

 \end{appendices}
  
\end{document}